\documentclass[preprint]{elsarticle}
\usepackage{amssymb}
\usepackage{graphicx}
\usepackage{dcolumn}
\usepackage{bm}

\begin{document}

\title{ Electronic structure and  magnetism of new scandium-based full Heusler compounds: Sc\raisebox{-.2ex}{\scriptsize 2}CoZ (Z=Si, Ge, Sn)}
\author{A. Birsan$^{1,2}$}
\address{$^1$National Institute of Materials Physics, 105 bis Atomistilor Street, PO Box MG.7, 077125 Magurele-Ilfov, Romania. \\
 $^2$ University of Bucharest, Faculty of Physics, 105 Atomistilor Street, PO Box MG-11, 077125, Magurele-Ilfov, Romania}

\begin{abstract}
 First principles FPLAPW calculations were performed in the framework of Density Functional Theory (DFT), to study the electronic structures and magnetic properties for the new full-Heusler compounds: Sc\raisebox{-.2ex}{\scriptsize 2}CoZ (Z=Si, Ge, Sn). The investigated materials are stable against decomposition, in ferromagnetic configuration and crystallize in the inverse Heusler structures. The half-metallic properties as function of the variation of unit cell volumes are  analysed regarding the fourth main group constituent elements. The electronic structure calculations for Sc\raisebox{-.2ex}{\scriptsize 2}CoSi and Sc\raisebox{-.2ex}{\scriptsize 2}CoSn show half-metallic characters, with indirect band gaps of 0.544 eV and 0.408 eV at optimised lattice parameters of 6.28  $\dot{A}$ and 6.62 $\dot{A}$, respectively. For Sc\raisebox{-.2ex}{\scriptsize 2}CoGe compound, the Fermi energy is not pinned inside the energy band gap from minority density of states, neither for unit cell contraction nor for enlargement. The calculated total magnetic moments are 1$\mu_{B}/f.u.$, for all compounds, in agreement with Slater-Pauling rule.

\end{abstract}


\maketitle

\section{Introduction}

Potential candidates for spintronic applications, the full-Heusler   compounds  were intensively studied to understand their physical properties \cite{Heusler1903A,Heusler1903B}. Nowadays, the interest in Heusler compounds grew enormously due to the discovery of materials with high Curie temperature, high spin polarization and low saturation of magnetization, which may be useful for new devices, related to the magnetic storage of information \cite{Graf2011}. Another breakthrough was the  perpendicular magnetic anisotropy, reported in thin films \cite{Wu2009} which may be incorporated in spin torque devices. Finding novel materials  with  designed properties suitable for tunnelling magnetoresistance (TMR) devices \cite{Kammerer2004, Herbot2009}, magnetic tunnel junctions \cite{Wurmehl2006,Wang2010,Sargolzaei2006} or spin injection devices \cite{Coey2002} is another active field of ongoing research involving Heusler compounds.

The Heusler family presents two possible variants of compounds: the half-Heusler compounds, $XYZ$, 1:1:1 stoichiometry, with non-centrosymmetric cubic structure $C1_{b}$ \cite{Heusler1903A,Heusler1903B,Webster1988}, and the full-Heusler compounds $X_{2}YZ$, 2:1:1 stoichiometry, that crystallize either in the $Cu_{2}MnAl$ ($L2_{1}$) structure with the cubic space group $Fm\bar{3}m$ \cite{Heusler1903A,Heusler1903B,Webster1988}, or in the $Hg_{2}CuTi$ prototype, known as "inverse Heusler structure"  with $F\bar{4}3m$ space group \cite{Kandpal2007}.

The inverse Heusler structure is found  in full-Heusler compounds when the Y element is more electronegative than X, all symmetries adopted are $ T_{d} $ and no position with $ O_{h} $ symmetry is present. In this case, the $X$ atoms are located in the non-equivalent 4a (0,0,0) and 4c (1/4,1/4,1/4) Wychoff positions, while the Y and Z atoms occupy the 4b (1/2,1/2,1/2) and 4d (3/4,3/4,3/4) positions, respectively.

The feature called half-metallic ferromagnetism, proposed in 1983 by Groot et al. \cite{deGroot1983} describes materials behaving as hybrids between metals and semiconductors or isolators \cite{Bradley1934}. We have used ab initio electronic-structure calculations to identify new half-metallic ferromagnetic materials, which are compatible with semiconductors, due to large enough spin-flip gaps.  Although many Heusler compounds have been intensively investigated experimentally or theoretically and predicted to have half-metallic properties \cite{Graf2011,Kammerer2004,Kandpal2007,Ouardi2013,MingYin2013,Gao2013,Endo2012}, $Sc_{2}Co$- based full-Heusler compounds with Z being an element from the fourth main group of the periodic table, have not been thoroughly  studied to  acquire an insight into the electronic structures and their magnetic properties. Therefore, the density of states, bandstructures and magnetic properties were investigated for the $Sc_{2}CoZ$ (Z=Si, Ge, Sn) full-Heusler compounds, based on Density Functional Theory. The results of theoretical investigations of the $Sc_{2}Co$- based full-Heusler compounds are analysed  in the present work. To the best of our knowledge, this is actually the first time when  $Sc_{2}CoGe$ and $Sc_{2}CoSn$ are reported up to now. The obtained theoretical results are compared with electronic structures and magnetic studies reported in literature for other  Heusler compounds.

\section{METHOD OF CALCULATION}
The half-metallic properties of $Sc_{2}CoZ$ (Z=Si, Ge, Sn) compounds have been subsequently examined via the self-consistent Full Potential Linearized Augmented Plane Wave (FPLAPW) method, implemented in WIEN2k code \cite{Wien}. The Perdew Burke Ernzerhof (PBE) with Generalized Gradient Approximation (GGA) \cite{Perdew1996A,Perdew1996B} was used for the exchange and correlation interaction. The muffin-tin radii ($R_{MT}$) chosen were  2.35 a.u. (Sc and Co), 2.17 a.u. (Si), 2.23 a.u. (Ge) and 2.42 a.u. (Sn), respectively. The selected energy threshold used to separate the core and valence states was -6 Ry. Within the modified tetrahedron method, a 46x46x46 mesh, containing 2456 irreducible k points was selected for the Brillouin zone (BZ) integration \cite{Blochl1994}.  The cut-off condition, employed to determine the number of plane waves was $K_{max}R_{MT} = 7$, ($K_{max}$ represents the maximum modulus of the reciprocal lattice vector). The convergence of the self consistent calculations was performed considering an integrated charge difference lower than $10^{-4}e/a.u.^{3}$ between two successive iterations and an energy convergence criterion no higher than $10^{-5}$ eV.

\section{RESULTS AND DISCUSSIONS}
\begin{figure}
 \begin{center}
    \includegraphics[scale=1]{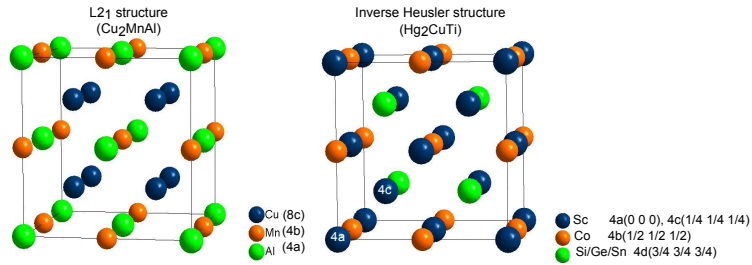} 
      \end{center}
   \caption{Possible prototype structures of crystallization for the full-Heusler compounds, the $L2_{1}$ ($Cu_{2}MnAl$) and inverse Heusler structure ($Hg_{2}CuTi$). The Wyckoff positions are illustrated in the figure.}
    \label{fig:structura}
\end{figure}

The most stable structures of the  $Sc_{2}CoZ$ (Z=Si, Ge, Sn) compounds were verified by performing structural optimizations (ferromagnetic configurations), with  the two possible prototype structures for crystallization of full-Heusler compounds, $Cu_{2}MnAl$ ($L2_{1}$)  and $Hg_{2}CuTi$ (the inverse Heusler structure), displayed in Fig.\ref{fig:structura}.

\begin{figure}
 \begin{center}
    \includegraphics[scale=0.8]{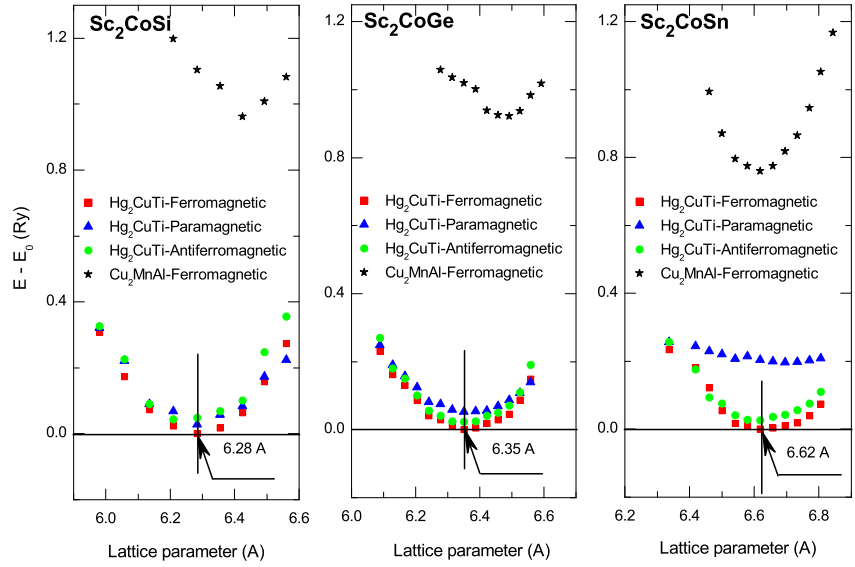} 
      \end{center}
   \caption{The calculated total energies as function of the lattice constants for $Sc_{2}CoZ$ (Z=Si, Ge, Sn)  for ferromagnetic, paramagnetic and antiferromagnetic configurations with $Hg_{2}CuTi$ -type structure and ferromagnetic configurations with $Cu_{2}MnAl$ prototype}
    \label{fig:optimizare}
\end{figure}

The calculations revealed the fact that the ordered $L2_{1}$ structure is unstable for all compounds, specifically, the model system with  $Cu_{2}MnAl$ prototype has higher energy compared to the system with $Hg_{2}CuTi$ -type structure (Fig.\ref{fig:optimizare}) and the latter prototype was used for further calculations. The result is in agreement with the occurrence of inverse Heusler structure reported in literature for $Ti_{2}$ -based full-Heusler compounds, \cite{Skaftouros2013, Feng2013,Birsan2013JMMM,Huang2012}, because Co atoms are more electronegative than Sc atoms.

 Moreover, the non-magnetic and antiferromagnetic configurations were examined and the  total energies calculated for these cases have higher values comparing with ferromagnetic configurations, at the equilibrium lattice parameters, for all the three considered compounds ( Fig.\ref{fig:optimizare}). 
The equilibrium lattice constants calculated for relaxed spin-polarized configurations are 6.28, 6.35 and 6.62 $\dot{A}$ for $Sc_{2}CoZ$ (Z=Si, Ge, Sn).  Recently reported \cite{Skaftouros2013}, the equilibrium lattice parameter for $Sc_{2}CoSi$,  (6.29 $\dot{A}$), is close to our result, while for the $Sc_{2}CoGe$  and $Sc_{2}CoSn$ compounds, there are no available lattice parameters reported in literature, to be compared with our calculated values.

For each full-Heusler compound ($Sc_{2}CoSi$, $Sc_{2}CoGe$ or $Sc_{2}CoSn$), the enthalpy change $\Delta{H}$ was calculated by subtracting the sum of equilibrium total energies for constituent elements ($Sc$ and $Co$ with HCP structures, $Si$, $Ge$ and $Sn$ with FCC structures), from the equilibrium total energies of corresponding compounds under study here, according to formula: 

$\Delta{H}=E_{Sc_{2}CoZ}-2E_{Sc}-E_{Co}-E_{Z}$, (Z=Si, Ge, Sn).

The values obtained for enthalpies of formation, in ferromagnetic configurations, at the calculated equilibrium lattice parameters are -0.51 eV/atom ($Sc_{2}CoSi$), -0.63 eV/atom ($Sc_{2}CoGe$) and -0.77 eV/atom ($Sc_{2}CoSn$)and reflect the stability of compounds against decomposition. 
\begin{figure}
 \begin{center}
    \includegraphics[scale=0.8]{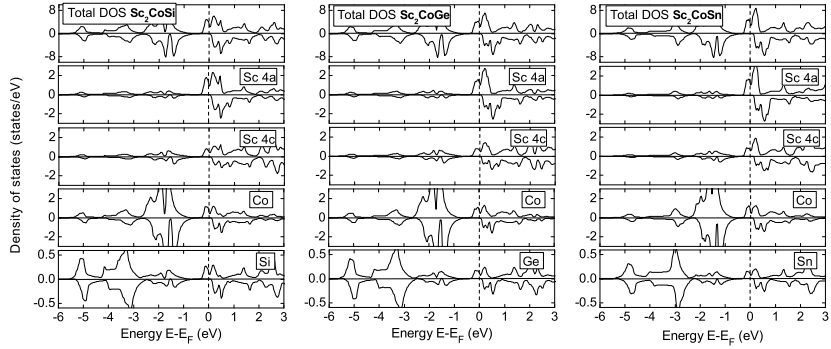} 
      \end{center}
   \caption{The spin-polarized total densities of states (DOS) and partial DOSs of $Sc_{2}CoZ$ (Z=Si, Ge, Sn) calculated at equilibrium lattice constants.}
    \label{fig:Dos-all}
\end{figure}

Fig. \ref{fig:Dos-all} displays the total and partial density of states calculated at equilibrium lattice parameters. In the majority spin channels, all compounds present a typical metallic behaviour. The minority spin channels of $Sc_{2}CoSi$ and $Sc_{2}CoSn$ present semiconducting characters with indirect band gaps around the Fermi levels for optimised structures. The metallic character from majority spin channels and the clear semiconducting gaps from minority spin channels lead to a full spin polarization of these compounds and a stable half-metallic behaviour at optimised lattice parameters. In the case of $Sc_{2}CoGe$, the Fermi level intersects the bottom of the conduction band. The pseudogap formed in the minority spin channel may be due to  very small effects, such as exchange and correlation effects, which can alter the spin polarization of this material by changing the density of states surrounding $E_{F}$. 
For all Heusler compounds analysed, in both spin channels, significant contributions to the total density of states in the energy range between -5.5 eV and -2.6 eV, come from  $p$ electrons of Z elements and $d$ electrons of Co atoms, the latter having the main contribution also between  -2.4 eV and -0.6 eV. 

\begin{figure}
 \begin{center}
    \includegraphics[scale=0.5]{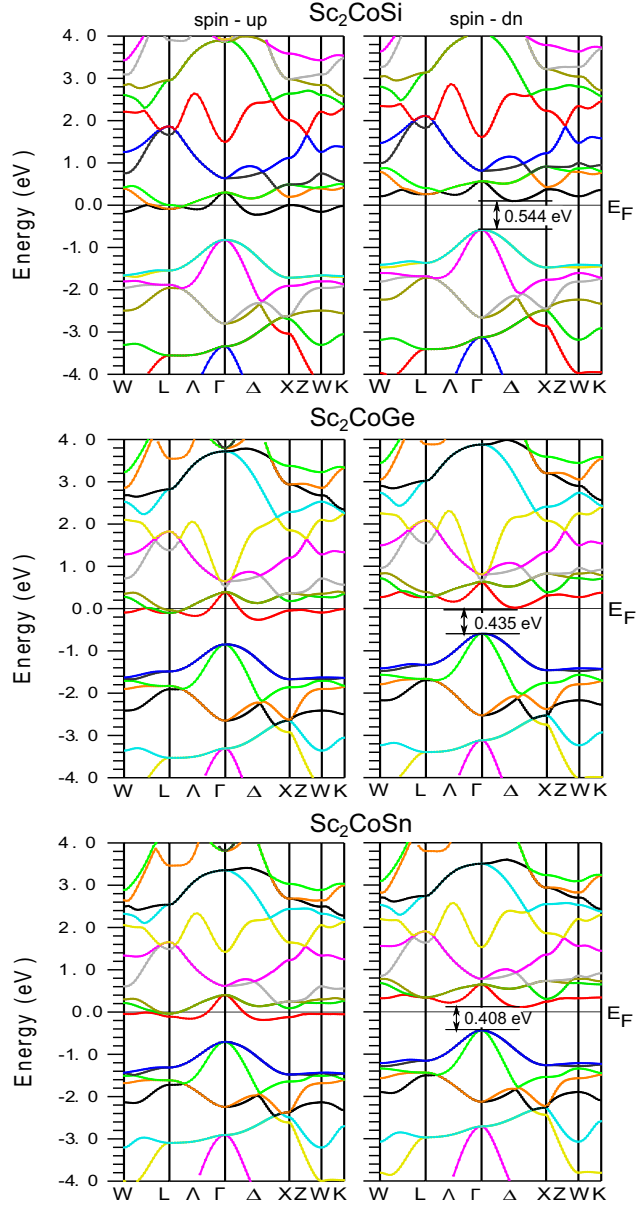} 
      \end{center}
   \caption{ (color) The bandstructures of $Sc_{2}CoZ$ (Z=Si, Ge, Sn) Heusler compounds. Majority spin channel (spin-up) in the left hand side of the figure, minority spin channel (spin-dn) in the right hand side.}
    \label{fig:benzi}
\end{figure}

The band structures of $Sc_{2}CoSi$, $Sc_{2}CoGe$ and $Sc_{2}CoSn$ compounds are displayed in Fig. \ref{fig:benzi}, at optimized lattice constants, with the majority spin channel on the left hand side of the figure and the minority spin channel on the right hand side. The size of the indirect band gaps, at optimized  lattice parameters, from the minority spin channels, are 0.544 eV for $Sc_{2}CoSi$ and 0.408 eV, for $Sc_{2}CoSn$ compound, respectively (Fig. \ref{fig:benzi}). Even though an indirect  band gap of 0.435 eV is present in the minority spin channel of $Sc_{2}CoGe$ compound, the Fermi level intersects the bottom of the conduction band, which leads to a loss of the semiconducting behavior and a decrease in the spin polarization at optimised lattice parameter.  The sizes of semiconducting gaps from minority spin channels, for all compounds, are determined by the difference in energy between the top of the valence bands, located at $\Gamma$ - point, below the $E_{F}$ and the bottom of the conduction bands, at $\Delta$ point, above the $E_{F}$. 

\begin{figure}
 \begin{center}
    \includegraphics[scale=0.8]{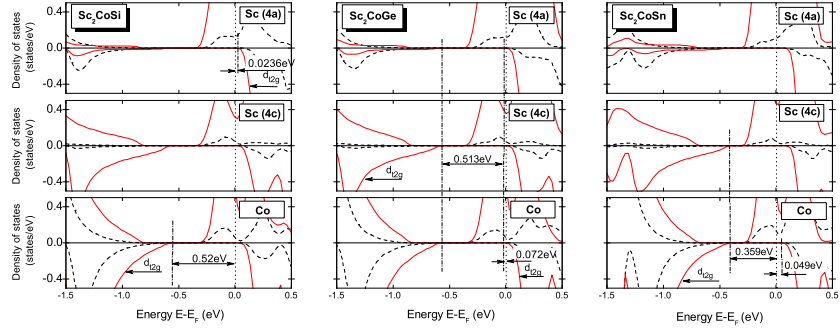} 
      \end{center}
   \caption{The main partial densities of states at optimized lattice parameters of
   $Sc_{2}CoZ$ (Z=Si, Ge, Sn), Fermi levels, $d_{eg}$ and $d_{t2g}$ being indicated by dotted, dashed and solid lines, respectively. }
    \label{fig:sitedos}
\end{figure}
The partial density of states of $d$-orbitals of Sc and Co atoms are illustrated in Fig. \ref{fig:sitedos} for all compounds, at optimised lattice parameter.
The band gaps from minority spin channels are formed by the d states of $Sc$(4c), $Sc$(4a) and $Co$ atoms, according to our theoretical calculations. In $Sc_{2}CoSi$ case, the origin of the band gap results from the coupling between the triple degenerated $d_{t2g}$ orbitals of $Sc$(4a) atoms, which  have energy values in the region corresponding to the anti-bonding states of the conduction band (0.0236 eV above $E_{F}$) and $d_{t2g}$ orbitals of $Co$ atoms that correspond to the bonding states of valence band (0.52 eV below $E_{F}$). The hybridization is dependent by the atomic arrangement of atoms and environment. It occurs when the sum of metallic radii (12-coordinated) of two neighbours exceeds the interatomic distance. Since the Pauling electronegativity of scandium is 1.3 and of Co is 1.9 (Pauling units), in $Sc_{2}CoSi$ case, the sum of metallic radii is higher than the distance between the next nearest neighbours Sc(4a)-Co, which is $a$/2, where $a$ is the lattice parameter. Therefore, Sc(4a) and Co atoms may strongly hybridize. The formation of the pseudogap in $Sc_{2}CoGe$ compound is explained by the interaction between the triple degenerated orbitals $d_{t2g}$ of $Co$  with contribution of $d_{t2g}$ orbitals of Sc(4c) at the bonding states of valence band, $Sc$ (4c) and $Co$ being the nearest neighbours. The Fermi level is estimated to be approximately 0.072 eV above the bottom of the conduction band. The indirect band gap from minority spin channel of $Sc_{2}CoSn$ is described by the hybridization  of $d_{t2g}$ anti-bonding  orbitals of $Co$ (0.049 eV above $E_{F}$) with the $d_{t2g}$ bonding  orbitals from $Co$ and $Sc$ (4c) atoms (0.359 eV below $E_{F}$).

\begin{figure}
 \begin{center}
    \includegraphics[scale=0.7]{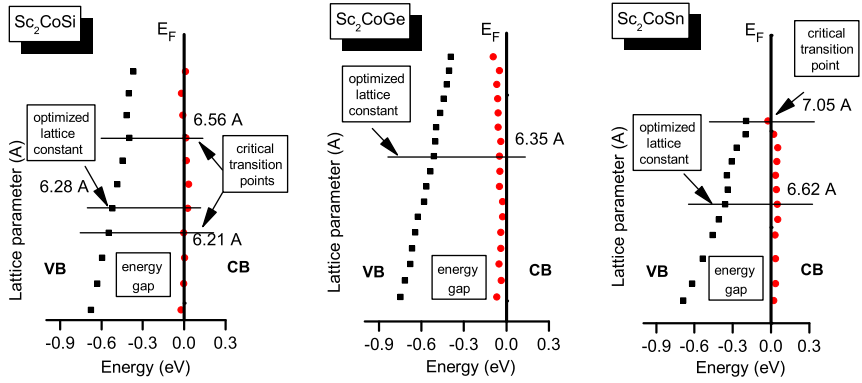} 
      \end{center}
   \caption{The positions of the top of the valence band (solid black squares) and the bottom edges of the conduction band (solid red circles) of total DOSs for $Sc_{2}CoZ$ (Z=Si, Ge, Sn) as function of lattice parameters, in minority spin channels.}
    \label{fig:gapSc2Co}
\end{figure}

Fig. \ref{fig:gapSc2Co} displays the half-metallic properties of for $Sc_{2}CoZ$ (Z=Si, Ge, Sn) as function of lattice parameters. In the case of $Sc_{2}CoSi$, the half-metallic band gap is present at the Fermi level, for a lattice parameter range 6.21 - 6.56 $\dot{A}$, providing a complete spin polarization. Below and above the critical transition points illustrated in Fig. \ref{fig:gapSc2Co}, the compound behaves as a conventional metal. The calculations for $Sc_{2}CoGe$  do not yield to half-metallic ferromagnetism. Because the Fermi level crosses the bottom edge of minority conduction band, the calculated spin polarization is $98\%$, at optimised lattice constant.
For each lattice parameter, the calculated Fermi energy is different; therefore, the Fermi level is shifted. However for $Sc_{2}CoGe$ compound, the Fermi energy is not pinned inside the energy band gap from minority density of states, neither for unit cell  contraction nor for enlargement. The $Sc_{2}CoSn$ compound shows a stable and well-ordered half-metallic ferromagnetism, being fully spin polarized, up to a lattice parameter of 7.05 $\dot{A}$. Above this transition point, the Fermi level intersects the bottom edge of the minority conduction band and the $Sc_{2}CoSn$ compound becomes ferromagnetic metal.

\begin{figure}
 \begin{center}
    \includegraphics[scale=0.8]{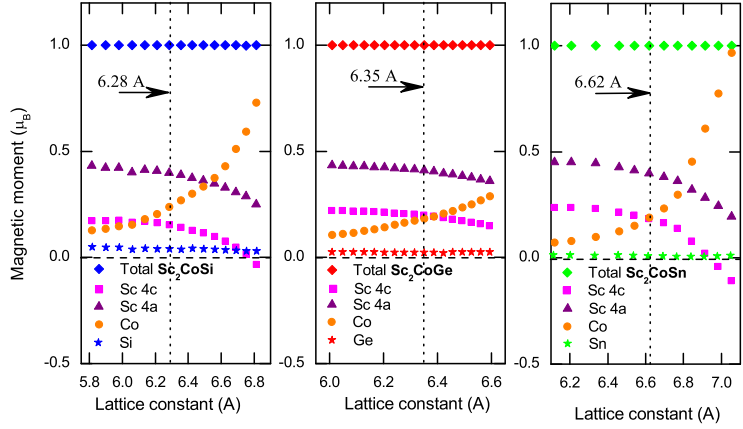} 
      \end{center}
   \caption{The total, site-projected magnetic moments as function of lattice constant for $Sc_{2}CoZ$ (Z=Si, Ge, Sn) compounds. }
    \label{fig:magnmom}
\end{figure} 
The Slater-Pauling curve gives in ferromagnetic alloys, the interrelation between the total magnetic moment and the valence electron concentration \cite{Slater1936,Pauling1938}. The original Slater-Pauling approach suggests the existence of different laws, due to the average over all atoms of the total magnetic moment and the number of valence electrons. For compounds with different kinds of atoms and ordered crystalline structures, it is more appropriate to consider all atoms of the unit cell, to find the magnetic moment per unit cell. For ternary 1:1:1 Heusler compounds, the Slater-Pauling rule was first time reported by K\"{u}bler \cite{Kubler1984}. These compounds, with $C1_{b}$ structure have three atoms per unit cell and follow the Slater-Pauling “18-electron-rule” ($M_{t}$ = $Z_{t}$ - 18), where $M_{t}$ is the total spin magnetic moment per the formula unit,  $Z_{t}$ is the total number of valence electrons and 18 represents the number of occupied states in the spin bands. A Slater-Pauling “24-electron-rule” ($M_{t}$ = $Z_{t}$ - 24) was found for the 2 :1 :1 family of  full-Heusler compounds with $L2_{1}$ structure ($Cu_{2}MnAl$ -prototype) \cite{Galanakis2002}. The present work deals only with ternary 2:1:1 full-Heusler compounds with $Hg_{2}CuTi$ – type structure.  Even though the origin of the band gap in the latter 2:1:1 full-Heusler compounds is different than that of the ternary 1:1:1 Heusler compounds, the corresponding Slater-Pauling rule is similar: “18-electron-rule” ($M_{t}$ = $Z_{t}$ - 18) . This Slater-Pauling “18-electron-rule” was  recently explained for $Ti_{2}$-based full-Heusler compounds \cite{Skaftouros2013,Birsan2013interm}. 
The total and site resolved magnetic moments as function of lattice constants for  $Sc_{2}CoZ$ (Z=Si, Ge, Sn) compounds are shown in Fig.\ref{fig:magnmom}. All materials studied have the total magnetic moments calculated at equilibrium lattice parameters  equal to 1 $\mu_{B}/f.u$ and follow the Slater-Pauling rule (“18-electron-rule”) for ternary 2:1:1 full-Heusler compounds with $Hg_{2}CuTi$ -type structure.   The calculated total and atomic spin magnetic moments are given in table I.

\begin{table}
 \begin{tabular}{|c|c|c|c|c|c|c|}
 \hline  & $a$  & $M_{t}$ & $m^{Sc(4a)}$ & $m^{Sc(4c)}$ & $m^{Co}$ & $m^{Z}$   \\ 
 \hline $Sc_{2}CoSi$ & 6.28 & 0.999 & 0.399 & 0.154 & 0.238 & 0.038  \\
        $Sc_{2}CoGe$ & 6.35 & 1.000 & 0.411 & 0.200 & 0.182 & 0.021  \\ 
        $Sc_{2}CoSn$ & 6.62 & 1.000 & 0.401 & 0.188 & 0.192 & 0.010  \\ 
 \hline 
 \end{tabular} 
 \caption{The optimised lattice parameters (in $\dot{A}$), total spin magnetic moments (in $\mu_{B}$), atomic spin magnetic moments ($\mu_{B}/atom$) calculated for the optimised lattice parameter, for $Sc_{2}CoZ$ (Z=Si, Ge, Sn). The differences between the calculated total magnetic moments and the  sum of atomic spin magnetic moments  represent the contributions of the interstitial regions.}
 \end{table}
 
For all compounds studied, the major contributions to the total magnetic moments come from Sc atoms, located in (4a) Wyckoff positions that have tetrahedral symmetries $T_{d}$ and are surrounded by  Co atoms, at $a/2\:\dot{A}$ distances ($a$ are the optimised lattice parameters).  This result is surprising because the Sc atoms in standard state do not have magnetic properties.  Similar findings were reported in $Ti_{2}$-based half-metallic full Heusler compounds, which also follow the “18-electron-rule”, where the  highest spin magnetic moment  contributions come from Ti atoms,  \cite{Birsan2013JMMM,Feng2011,Huang2012}. 

Despite the fact that Sc atoms are coupled ferromagnetically, their different neighbourhoods  determine dissimilar magnetic moments. Although the main group elements ($Si$, $Ge$ or $Sn$) carry an insignificant  magnetic moment, the Z element surroundings influence the size of the band gap, from minority spin channel, by decreasing it, as the atomic radius of the main element becomes larger. Furthermore, the atomic spin magnetic moments of Sc atoms from both sites are decreasing, regardless of the main group elements.  
In the materials containing Si or Sn, the stress applied to structures, leading to unit cell expansions, determines the Co magnetic moments to increase. In the case of $Sc_{2}CoGe$ compound, the magnetic moment of Co slightly increases, but does not exceed the magnetic moment of Sc (4a), in spite of unit cell enlargement. 

However, the highest spin magnetic moments from $Sc_{2}$-based compounds are material specific. For example, in $Sc_{2}YZ$, (Y = Cr, Mn; Z = Al, Si) compounds, the highest magnetic moments come from transition metal elements located in Y position and coupled antiferromagnetically with Sc atoms from both Wyckoff positions \cite{Skaftouros2013}.   

\section{Conclusions}
To summarise, in the present study, first principles (FPLAPW) calculations investigate the half-metallic features of $Sc_{2}CoSi$, $Sc_{2}CoGe$ and $Sc_{2}CoSn$ full-Heusler compounds, with $Hg_{2}CuTi$ -type structure,  as potentially suitable materials for applications in spintronics. The electronic structure calculations and magnetic properties predict that $Sc_{2}CoSi$ compound is fully spin polarised half-metallic ferromagnet for lattice parameters ranging between 6.21 - 6.56 $\dot{A}$. The size of the calculated band gap, from minority density of states is 0.544 eV, at optimised lattice parameter of 6.28 $\dot{A}$. For $Sc_{2}CoGe$ compound, at the equilibrium lattice parameter 6.35 $\dot{A}$, is expected a reduction in the spin polarization at 98$\%$, due to the formation of a pseudogap near Fermi level. 
The band structure of $Sc_{2}CoSn$  is completely spin polarized, with a semiconducting indirect band gap in minority spin channel at optimized lattice parameter, 6.62$\dot{A}$. The transition from half-metallic to metallic features of $Sc_{2}CoSn$ occurs at a very enlarged lattice parameter of 7.05 $\dot{A}$, which makes the full-Heusler compound, the most stable half-metallic ferromagnet between the $Sc_{2}CoZ$ (Z=Si, Ge, Sn) studied compounds .

\section{ACKNOWLEDGMENTS}
The author thanks Dr. P. Palade for his support, and Dr. V. Kuncser for helpful discussions. This work was financially supported from the project  PNII IDEI 75/2011 of the Romanian Ministry of Education Research, Youth and Sport. 

\bibliographystyle{elsarticle-harv}

\end{document}